\documentclass[12pt]{article}
\usepackage{amsmath}
\usepackage{amsfonts}
\usepackage{amssymb}
\usepackage{amsthm}
\usepackage{cite}
\usepackage{hyperref}
\usepackage{mathrsfs}
\usepackage{subcaption}
\usepackage{cancel}
\usepackage{color}  
\newcommand{\bc}{\color{black}}

\newlength{\dinwidth}
\newlength{\dinmargin}

\usepackage{tikz}
\usetikzlibrary{trees}
\usetikzlibrary{decorations.pathmorphing}
\usetikzlibrary{decorations.markings}

\newcommand{\mrm}{\mathrm}

\renewcommand{\mathbf}{\boldsymbol}

\newcommand{\mcR}{\mathcal R}
\newcommand{\mcF}{\mathcal F}

\newcommand{\mcL}{\mathcal L}

\renewcommand{\i}{\mathrm i}

\newcommand{\wt}{\widetilde}

\newcommand{\ti}{\tilde}

\newcommand{\Om}{\Omega}

\newcommand{\ka}{\kappa}

\newcommand{\vk}{k}

\newcommand{\wh}{\widehat}

\newcommand{\ov}{\overline}

\newcommand{\eps}{\varepsilon}

\newcommand{\e}{\mathrm{e}}

\newcommand{\nin}{\noindent}
\newcommand{\si}{\sigma}

\newcommand{\h}{\fr{1}{2}}
\newcommand{\nat}{\mathbb{N}}
\newcommand{\hil}{\mathcal{H}}

\newcommand{\mfa}{\mathfrak{A}}
\newcommand{\mco}{\mathcal{O}}

\newcommand{\supp}{\mathrm{supp}}
\newcommand{\fr}[2]{\frac{#1}{#2}}
\newcommand{\al}{\alpha}
\newcommand{\real}{\mathbb{R}}
\newcommand{\complex}{\mathbb{C}}

\newcommand{\non}{\nonumber}

\newcommand{\lan}{\langle}
\newcommand{\ran}{\rangle}

\def\proof{\noindent{\bf Proof. }}
\def\qed{$\Box$\medskip}

\newtheorem{theoreme}{Theorem } [section]
\newtheorem{proposition}[theoreme]{Proposition}
\newtheorem{lemma}[theoreme]{Lemma}
\newtheorem{definition}[theoreme]{Definition}
\newtheorem{corollary}[theoreme]{Corollary}
\newtheorem{remark}[theoreme]{Remark}
\newtheorem{example}[theoreme]{Example}
\newtheorem{criterion}[theoreme]{Criterion}
\newtheorem{conjecture}{Conjecture}
\newtheorem{assumption}{Assumption}

\newcommand{\bea}{\begin{assumption}}
	\newcommand{\eea}{\end{assumption}}
\newcommand{\beco}{\begin{conjecture} }
	\newcommand{\eeco}{\end{conjecture} }
\newcommand{\beq}{\begin{equation}}
	\newcommand{\eeq}{\end{equation}}
\newcommand{\beqa}{\begin{eqnarray}}
	\newcommand{\eeqa}{\end{eqnarray}}
\newcommand{\ben}{\begin{arabicenumerate}}
	\newcommand{\een}{\end{arabicenumerate}}
\newcommand{\bex}{\begin{example}}
	\newcommand{\eex}{\end{example}}
\newcommand{\ber}{\begin{remark}}
	\newcommand{\eer}{\end{remark}}
\newcommand{\bec}{\begin{corollary}}
	\newcommand{\eec}{\end{corollary}}
\newcommand{\bep}{\begin{proposition}}
	\newcommand{\eep}{\end{proposition}}
\newcommand{\becr}{\begin{criterion}}
	\newcommand{\eecr}{\end{criterion}}

\newcommand{\1}{\!\!}

\def\bel{\begin{lemma}}
	\def\eel{\end{lemma}}
\def\bet{\begin{theoreme}}
	\def\eet{\end{theoreme}}
\def\bed{\begin{definition}}
	\def\eed{\end{definition}}

\begin{document}

\title{Local normality of infravacua and relative normalizers for relativistic systems} 
\author{Bartosz Biadasiewicz and Wojciech Dybalski\\\\
Faculty of Mathematics and Computer Science \\  
Adam Mickiewicz University in Pozna\'n\\
ul. Uniwersytetu Pozna\'nskiego 4, 61--614 Pozna\'n, Poland.\\
\small{E-mails: {\tt bartosz.biadasiewicz@amu.edu.pl,  wojciech.dybalski@amu.edu.pl. }} \\\\
}
\date{}
\maketitle
\begin{abstract}  We revisit the problem of local normality of Kraus-Polley-Reents 
infravacuum representations and provide a straightforward proof based on the Araki-Yamagami
criterion. We apply this result to the   theory of superselection sectors.
Namely, we extend the novel formalism of second conjugate classes
and relative normalizers  to the local relativistic setting.

\end{abstract}

\section{Introduction}
\setcounter{equation}{0}

The infrared problem in QFT is a maze of difficulties caused by massless particles. 
In the algebraic approach  one aspect of this problems is a multitude
of superselection sectors differing by  \emph{soft photon clouds}  which escape detection. 
It is therefore natural to group these sectors into equivalence classes   and several 
definitions of such \emph{charge classes} are available in the literature \cite{Bu82,BR14, CD18}.
The most recent approach from \cite{CD18}, based on a novel concept of the relative
normalizer (see formula~(\ref{N}) below), concerns the structure of the group of automorphisms $\mrm{Aut}(\mfa)$
of the $C^*$-algebra $\mfa$ of observables. A restrictive aspect of this group theoretic
approach is that all the relevant representations of $\mfa$ must be expressible as automorphisms
in the defining (`vacuum') representation. The conventional Kraus-Polley-Reents (KPR) infravacuum representation \cite{Re74, KPR77, Kr82},
which describes a background radiation blurring the soft photon clouds,  is immediately expressible by an automorphism in a  non-relativistic case considered in \cite{CD18}. However, in the relativistic setting
it is not obvious if an associated automorphism exists and one may wonder if the group theoretic formalism
of \cite{CD18} generalizes to this context.   As we show in Section~\ref{local-system}, using a result of Takesaki 
\cite{Ta70}, this is actually the case, provided that the infravacuum representation is \emph{locally normal}. 
 
For us this by itself  is sufficient motivation to revisit the problem of local normality of KPR  representations
of the massless scalar free field. This property is actually claimed by Kunhardt in  \cite[Proposition 3.4]{Ku98}, but only some hints for the proof are given with a reference for details to an unpublished work of F.~Hars.  {\bc However, we are not going  to reconstruct the strategy indicated in this reference as  it is based on the phase space condition $C_{\sharp}$ \cite{BP90}}. Firstly, to our knowledge, this condition has only been verified  for free scalar fields but not
e.g. for free electromagnetism. This would suffice for the present paper, but not for planned generalizations. 
Secondly, and more importantly, condition $C_{\sharp}$ is not expected to hold for unbounded regions such us, e.g., 
future lightcones.  Yet we consider  \emph{lightcone normality} of infravacuum representations an important question 
for future research. It is relevant, in particular, for exemplifying the abstract constructions of Buchholz and Roberts  from \cite{BR14}. {\bc We remark that a much simpler question of lightcone normality of coherent states has been settled
only recently in \cite{CD20}.}

In this paper we aim for a more optimal strategy for proving the local normality of KPR representations. 
{\bc This} question  is  related to the well-known Shale-Stinespring problem of unitary implementation of
symplectic transformations on Fock space \cite{Sh62,SS65, Ru78}, which is nowadays textbook material \cite{Ar, DG1,HSSS12}.  However, since we  aim for local and not global normality (the latter is actually in conflict with the infravacuum property (\ref{S})), the symplectic
form is effectively degenerate, which excludes the above formulations. A  Shale-Stinespring type
theorem valid in the degenerate  case was proven by Araki and Yamagami \cite{AY82} and we will rely on this result here.       
Actually, the same route was taken in several other investigations of local normality in scalar free field theory on flat and 
curved spacetime, e.g.  for Hadamard states \cite{Ve94} and for certain infravacua in the two-dimensional massless case \cite{BFR21}.  

Our paper is organized as follows: In Section~\ref{local-system} we demonstrate that local normality 
allows to generalize the formalism of relative normalizers and second conjugate classes from \cite{CD18}
to the relativistic framework. In Section~\ref{Preliminaries}  we describe a class of quasi-free representations
of the  massless scalar field, which are given by symplectic transformations $T$.
We list conditions on $T$ which imply the irreducibility and infravacuum property of these representations. 
In Section~\ref{Araki-Yamagami-section} we use a result of Araki and Yamagami \cite{AY82} to formulate conditions on $T$ which guarantee local normality of the resulting representation. These conditions are verified in 
Section~\ref{KPR-maps}  in the case of the KPR infravacuum map. Given observations from Section~\ref{local-system} 
this concludes a construction of a non-trivial relative normalizer in the local relativistic case.

\vspace{0.2cm}

\noindent{\bf Acknowledgments:} W.D. would like to thank D. Buchholz and S. Doplicher for helpful 
discussions on the literature. W.D. was partially supported by the  Emmy Noether grant  DY107/2-2  and  the NCN grant `Sonata Bis' 2019/34/E/ST1/00053.

\section{Relative normalizers for relativistic systems} \label{local-system}
\setcounter{equation}{0}

We focus here on the most recent  approach to building equivalence classes of sectors \cite{CD18}, which 
can be explained in very general terms: Let $\mfa$ be a $C^*$-algebra and $G:=\mrm{Aut}(\mfa)$ its automorphism group.
$G$ acts on the set of sectors $X$, i.e., orbits of pure states under the action of  
the group of inner automorphisms. Given a distinguished vacuum sector $x_0\in X$, 
the  \emph{second conjugate class} of $x=x_0\cdot g_x$,  $g_x\in G$, w.r.t. a \emph{background} $a\in G$
 is given by
\beqa
\ov{\ov{[x]}}^a:=[x]_{{\bc a^{-1}}\cdot G_{x_0}\cdot {\bc a}}, \label{second-conjugate}
\eeqa     
where $G_{x_0}$ is the stabilizer group of $x_0$ and the r.h.s. of (\ref{second-conjugate}) denotes the orbit
of $x$ under $a^{-1}\cdot G_{x_0} \cdot a$.  As discussed in \cite{CD18}, this definition is
motivated by conventional superselection theory, where the conjugation is involutive.

The soft photon clouds in this setting are
sectors of the form $x_0\cdot s$, $s\in S$, where the subgroup  $S\subset G$ is not contained
in $G_{x_0}$, that is, $x_0\cdot s\neq x_0$ for some $s\in S$. The second conjugate class (\ref{second-conjugate})
serves its purpose, if the background $a$ is chosen {\bc in such a way, that}
\beqa
\ov{\ov{[x_0\cdot s]}}^a= \ov{\ov{[x_0]}}^a. \label{cloud-absorption}
\eeqa  
A convenient sufficient condition is that  $a$ is an element of the \emph{relative normalizer}~\cite{CD18}
\beqa
N_{G}(R,S):=\{\, g\in G\,|\, g\cdot S\cdot g^{-1}\subset R\,\},  \label{N}
\eeqa
where $R\subset G_{x_0}$ is a subgroup. (We drop here the assumption $R\subset S$ from \cite{CD18}
as it is not needed for relation~(\ref{cloud-absorption})).

A search for suitable backgrounds, i.e., elements of $N_{G}(R,S)$, naturally leads to \emph{infravacuum representations}
$\pi: \mfa\to B(\hil)$. By definition, they satisfy
\beqa
\pi\cdot s\simeq \pi, \quad s\in S, \label{S}
\eeqa
where $\simeq$ denotes unitary equivalence. In the non-relativistic setting of \cite{CD18} the  KPR
infravacuum representation has the form $\pi=\pi_{\mrm{id}}\circ \al$, where $\pi_{\mrm{id}}$ is the defining vacuum representation and 
$\al$ is an automorphism of $\mfa$. From this and (\ref{S}) we immediately get $\al\cdot s\cdot \al^{-1}=\mrm{Ad}(U)$
for some unitary $U$ on the vacuum Hilbert space, hence $\al$ belongs to the relative normalizer (\ref{N}).  

As mentioned in the Introduction, in the relativistic setting the KPR infravacuum representations are
not immediately expressible by  automorphisms. 
However, it turns out that representations of local nets of von Neumann algebras are closely
related to automorphisms provided that they are \emph{locally normal}. This is a  content of a theorem by
Takesaki  \cite[Theorem 12]{Ta70}, which we now recall in a form adapted to our situation:
 {\bc Let $\mfa$ be the global $C^*$-algebra of a  net $\mco\mapsto \mfa(\mco)$ of infinite dimensional von Neumann algebras, labelled by open, bounded regions $\mco\subset \real^4$, satisfying isotony and locality. In addition, we assume the split property, that is for any open, bounded region $\mco_1$ there is another open, bounded region $\mco_2$ and 
 a type I factor  $\mcR$ s.t. 
 \beqa
 \mfa(\mco_1)\subset \mcR\subset \mfa(\mco_2).    \label{R}
 \eeqa
 Given such structure,  we say that a representation $\pi$ of $\mfa$ is \textbf{locally normal} if it is $\si$-weakly continuous on each local subalgebra $\mfa(\mco)$. There holds the following:
\bet\emph{\cite[Corollary 13]{Ta70}}\label{Takesaki} Let $\mfa$ be as  above. Suppose its defining representation $\pi_{\mrm{id}}$ acts irreducibly on a separable Hilbert space. Let $\pi$ be an  irreducible, locally normal representation of $\mfa$ on a separable Hilbert space. Then there exists an automorphism $\al$~s.t.
\beqa
\pi_{\mrm{id}}\cdot \al\simeq \pi. \label{Takesaki-alpha}
\eeqa
\eet
 We remark that the proper sequential funnel of type $I_{\infty}$ factors in $\mfa$, assumed in \cite{Ta70}, is readily
constructed from the factors $\mcR$ in (\ref{R}). The properness condition \cite[Definition 6]{Ta70} is verified 
using isotony and locality as well as the fact that the  relative commutant of type I factors is type I. This follows from
\cite[p.300]{Ta} and the fact that a type $I_{\infty}$ factor is quasi-equivalent to $B(\hil)$.

}
Now suppose that $\pi$ from Theorem~\ref{Takesaki} satisfies in addition the infravacuum property (\ref{S}).
Then we have for some unitaries $U, U', U''$ on $\hil$
\beqa
\pi_{\mrm{id}}\cdot \al\cdot s =\mrm{Ad}(U)\cdot  \pi_{\mrm{id}} \cdot s= \mrm{Ad}(U')\cdot  \pi =\mrm{Ad}(U'')\cdot \pi_{\mrm{id}}\cdot \al.
\eeqa
Since $\pi_{\mrm{id}}$ is the defining representation, this means $\al\cdot s\cdot \al^{-1}=\mrm{Ad}(U'')$. Thus we obtain:
\bec Let $\pi$, $\pi_{\mrm{id}}$ be as in theorem~(\ref{Takesaki}) and, in addition, $\pi$ be an infravacuum 
representation  in the sense of (\ref{S}). Then the automorphism $\al$ of (\ref{Takesaki-alpha}) 
is an element of the relative normalizer $N_{G}(R,S)$ of (\ref{N}).
\eec

\section{Symplectic maps and quasi-free representations} \label{Preliminaries}
\setcounter{equation}{0}

Let  us introduce a vector space $L:=D(\real^3;\complex) \oplus D(\real^3;\complex)$, whose elements are pairs of functions $G=(G_1,G_2)$. We equip it with the
symplectic form 
\beqa
\si(G, G')=  \int_{\real^3} (\ov{G}_1G_2'-\ov{G}_2G_1') dx.   \label{symplectic}
\eeqa
Now let  $\mu(k):=|k|$ and consider the vector spaces 
$\mcL_1:= \mu^{-1/2} \wh{D}(\real^3;\complex)$, $\mcL_2:=\mu^{1/2} \wh{D}(\real^3;\complex)$, where hat denotes the
Fourier transform. We denote elements of
$\mcL:=\mcL_1\oplus \mcL_2$ by $F=(F_1,F_2)$ and define a symplectic form on $\mcL$ by extending $\si$
to $L^2(\real^3)\oplus L^2(\real^3)$. We note that $\si(F,F')=\si(G,G')$  thus the mapping
\beqa
L\ni (G_1,G_2)\overset{F}{\mapsto} (\mu^{-1/2}G_1, \mu^{1/2}G_2)\in \mcL,    \label{symplectomorphism}
\eeqa
preserves the symplectic form. The subspaces of $L$ and $\mcL$, determined by $G_1, G_2$ supported
in a ball $O_r$ of radius $r>0$, centered at zero, will be denoted by $L_r$, $\mcL_r$.

Now define two complex-linear maps
\beqa
T_1: \mcL_1 \to L^2(\real^3), \quad  T_2:  \mcL_2 \to L^2(\real^3)\quad  \textrm{ s.t. }\quad  \lan T_1 F_1, T_2 F_2\ran=\lan F_1, F_2\ran, \label{T-def}
\eeqa
where $\lan \,\cdot\,,\,\cdot\,\ran$ is the scalar product in $L^2(\real^3)$.
Consequently, $T :\mcL\mapsto L^2(\real^3)\oplus L^2(\real^3)$ given by $T(F_1, F_2)=(T_1F_1, T_2 F_2)$
is a symplectic map.  {\bc We also require, that $T_1,T_2$ commute with complex conjugation in configuration space, as
this will be needed in (\ref{coherent-state-computation}) below.}

{\bc Now we impose the infravacuum property on this map. We introduce the following subspace of the
algebraic dual $\mcL^*$ of $\mcL$:
\beqa
\mcL^*_{S}:=\mu^{-3/2}\chi(\mu)C_{\mrm{sym}}^{\infty}(S^2),
\eeqa
where $\chi$ is the sharp characteristic function of some fixed interval containing zero and  $C_{\mrm{sym}}^{\infty}(S^2)$ denotes smooth, real-valued functions on the sphere, symmetric under 
$\hat{k}\mapsto -\hat{k}$. (Due to this later property, these functions are invariant under complex conjugation in configuration space).} 
{\bc We say that the map $T$ has the  \textbf{infravacuum property w.r.t. $\mcL^*_{S}$} if for any $\mrm{v}\in \mcL^*_{S}$
there exists an element of $L^2(\real^3)$, which we denote $T_1\mrm{v}$, s.t. 
\beqa
\lan \mrm{v}, F_2\ran=\lan T_1\mrm{v}, T_2F_2\ran  \textrm{ for all } F_2\in \mcL_2. \label{infravacuum-prop}
\eeqa
We note that the $L^2$-pairing on the l.h.s. of (\ref{infravacuum-prop}) is well defined 
 and (\ref{infravacuum-prop}) extends relation~(\ref{T-def}).}

Now let $\mcF$ be the symmetric Fock space and denote the usual creation and annihilation operators
by $a^*$, $a$ and the Fock space vacuum by $\Om$.    For any $G=(G_1,G_2)$ consider the scalar quantum field and canonical momentum in a representation specified by $T$: 
\beqa
 \phi_{T}(G_1)\1&:=&\1\fr{1}{\sqrt{2}} \big( a^*(T_1\mu^{-1/2} \wh{G}_1 \big)+ a(T_1\mu^{-1/2} \wh{\ov{G}}_1) \big), \\
\pi_{T}(G_2)\1&:=&\1\fr{1}{\sqrt{2}} \big( a^*(\i T_2\mu^{1/2} \wh{G}_2 \big)+ a(\i T_2\mu^{1/2} \wh{\ov{G}}_2) \big), \\
\Phi_{T}(G)\1&:=&\1\phi_T(G_1)+\pi_T(G_2).
 \eeqa
The case $T=\mrm{id}$, which reproduces the usual (vacuum) representation  will be indicated by dropping the index $T$.  We  introduce the local von Neumann algebra, corresponding to a double cone $\mco_r$, whose base is
the ball $O_r$,  
\beqa
\mfa(\mco_r):=\{\, \e^{\i \Phi(G) } \,|\, G \textrm{ real-valued, } \supp(G)\subset O_r\}'',
\eeqa
and the global $C^*$-algebra $\mfa:=\ov{\bigcup_{r>0} \mfa(\mco_r)}$. The algebras $\mfa(\mco)$, corresponding
to arbitrary open bounded regions $\mco\subset \real^4$ are now obtained in a standard manner \cite{Bo00}. 
It is well known that this net of algebras satisfies properties listed above Theorem~\ref{Takesaki}, in particular the split property \cite{BW86, BJ87}. 

We consider a representation $\pi_{T}: \mfa\to B(\mcF)$ defined by
\beqa
\pi_T(\e^{\i\Phi(G)})=\e^{\i \Phi_{T}(G) }, \quad G\in D(\real^3;\real)\oplus D(\real^3;\real).
\eeqa
We recall that $\pi_T$ is \textbf{irreducible} if
\beqa
\ov{\{ T_1F(G)_1+\i T_2F(G)_2\,|\, G\in D(\real^3;\real)\oplus D(\real^3;\real)\}}=L^2(\real^3), \label{irreducibility}
\eeqa
where $F$ is defined in (\ref{symplectomorphism}) \cite[Section 3.1]{Ku98}.  To state the infravacuum 
property for these representations,   we introduce the coherent automorphisms of $\mfa$ by extending
the relation
\beqa 
\al_{\mrm{v}}(\e^{\i\Phi(G)})=\e^{-\i\si(    (\mrm{v},0), F(G))}  \e^{\i\Phi(G)}, \quad G\in D(\real^3;\real)\oplus D(\real^3;\real),
\label{coherent-automorphism}
\eeqa
for $\mrm{v}\in \mcL^*_S$. {\bc (Here we could write $\si(    (\mrm{v},0), F(G))=\lan \mrm{v}, F(G)_2\ran$, by analogy with (\ref{symplectic}), since the $L^2$-pairing between elements of $\mcL^*_{S}$  and $\mcL_2$ is well defined).} We note a simple lemma which is implicit in \cite{Ku98}:
\bel\label{coherent-automorphisms}  Suppose that $T$ has the infravacuum property w.r.t. $\mcL^*_S$. Then $\pi_{T}$
has the infravacuum property w.r.t. $S:=\{\, \al_{\mrm{v}}\,|\, \mrm{v}\in  \mcL^*_S\}$, i.e.,
 \beqa
\pi_{T}\cdot  s\simeq \pi_T, \quad s\in S, \label{infravacuum-two}
\eeqa  
where $\simeq$ denotes the unitary equivalence. Furthermore, $\pi_T$ is not unitarily equivalent to
the defining representation $\pi_{\mrm{id}}$.
\eel
\proof For  $\mrm{v}\in \mcL^*_{S}$ the automorphism $\al_{\mrm{v}}$ is defined as  in (\ref{coherent-automorphism}).
We have, by the  infravacuum property of $T$,
\beqa
\pi_{T}\circ \al_{\mrm{v}}=\mrm{Ad}U_{\mrm{v}}\circ \pi_{T}, \label{infravacuum-property}
\eeqa
where $U_{\mrm{v}}:=\e^{\fr{\i}{\sqrt{2}}(a^*(T_1\mrm{v})+  a(T_1\mrm{v})  ) }$ is a unitary on $\mcF$.  This follows from the computation
\beqa
\pi_{T}\circ \al_{\mrm{v}}( \e^{\i\Phi(G)})=\e^{-\i\si( T(\mrm{v},0), TF(G))} \e^{\i\Phi_T(G)}=
U_{\mrm{v}}\pi_{T}(\e^{\i\Phi(G)} )   U_{\mrm{v}}^*, \label{coherent-state-computation}
\eeqa
{\bc which uses the CCR and the infravacuum property of $T$ defined in~(\ref{infravacuum-prop}).}
Now suppose that $\pi_{\mrm{id}}=\mrm{Ad}U\circ \pi_T$ for some unitary $U$.  Then, by (\ref{infravacuum-property}),
\beqa
\pi_{\mrm{id}}\circ \al_{\mrm{v}} =\mrm{Ad}U\circ \pi_T \circ \al_{\mrm{v}}
=\mrm{Ad}(UU_{\mrm{v}})\circ  \pi_T=\mrm{Ad}(UU_{\mrm{v}}U^*)\circ  \pi_{\mrm{id}}
\eeqa
This is a contradiction, since  $\pi_{\mrm{id}}\circ \al_{\mrm{v}}$,
 is disjoint from $\pi_{\mrm{id}}$ for some non-zero $\mrm{v}$ (cf. e.g. \cite{Ku98, CD18}). \qed
\section{Local normality of quasi-free representations} \label{Araki-Yamagami-section}
\setcounter{equation}{0}

{\bc The map $T$ in this section satisfies relation~(\ref{T-def}) and commutes with complex conjugation in configuration space. We do not require here  the infravacuum~(\ref{infravacuum-prop}) or the
irreducibility property~(\ref{irreducibility}). We will justify the following criterion for local normality:} 
\bet\label{main-result}   Fix $r>0$ and let $\chi_r\in D(\real^3;\real)$ be an approximate
characteristic function\footnote{$\chi_r$ should be equal to one on $O_r$ and vanish outside of a slightly larger set.} of $O_r$. Define operators $\chi_{1,r}:=\mu^{-1/2}\chi_r \mu^{1/2}$ and
$\chi_{2,r}:= \mu^{1/2}\chi_r \mu^{-1/2}$, which are bounded by Lemma~\ref{standard}. 
 Suppose that the following conditions hold:
\begin{enumerate}

\item $T_j\chi_{j,r}$ extend from $\mcL_j$ to bounded operators on $L^2(\real^3)$ and there exists  $c_r>0$ s.t. 
\beqa
  c_r  \chi_{j,r}^*\chi_{j,r}  \leq \chi_{j,r}^* (T_j^*T_j) \chi_{j,r}\leq c_r^{-1}  \chi_{j,r}^*\chi_{j,r}, \quad j=1,2. 
\eeqa

\item The following operators are trace class on $L^2(\real^3)$ 
\beqa
K_{1,r}:=\chi_r \mu^{-1/2}(T_1^*T_1-1)\mu^{-1/2}\chi_r,  \quad K_{2,r}:=\chi_r \mu^{1/2}(T_2^*T_2-1)\mu^{1/2}\chi_r.
\label{K-operators}
\eeqa

\end{enumerate}
Then $\pi_T$ is $\si$-weakly continuous on $\mfa(\mco_r)$. 
\eet
\nin We will prove this theorem  using  a criterion for quasi-equivalence of representations of CCR-algebras due to 
Araki and Yamagami \cite{AY82}. Thus we define a sesquilinear form on $L$
\beqa
S_{T}(G,G')\1&:=& \1 \lan \Om, \Phi_{T}(G)^* \Phi_T(G')\Om\ran, 
\eeqa
which for real-valued $G$ satisfies  $\lan \Om, \e^{\i\Phi(G) } \Om\ran=\e^{-\h S_{T}(G,G) }$, in accordance with  
\cite[Proposition 3.4 (iii)]{AY82}.  We observe, by explicit computations, that condition (1.3) of \cite{AY82} holds true
\footnote{The origin of the imaginary unit on the r.h.s. can be seen by  comparing our Weyl relations 
$\e^{\i \Phi_{T}(G) } \e^{\i \Phi_{T}(G') }=\e^{-\fr{\i}{2} \si(G,G') }  \e^{\i \Phi_{T}(G+G') }$ with 
\cite[Proposition 3.4 (ii)]{AY82}.}:
\beqa 
S_{T}(G,G)\geq 0, \quad S_{T}(G,G')-S_{T}(\ov{G}', \ov{G})=\i\si(G,G').
\eeqa
Next, we define the sesquilinear form
\beqa
( G | G')_{T}\1&:=&\1  S_{T}(G,G')+S_{T}(\ov{G}',\ov{G})\non\\
\1&=&\1    \big( \lan G_1, \mu^{-1/2}(T_1^*T_1)\mu^{-1/2} G_1'\ran+ 
\lan G_2, \mu^{1/2}(T_2^*T_2)\mu^{1/2} G_2'\ran\big) \label{scalar-product}
\eeqa
and note the following fact:
\bel The sesquilinear form $(\,\cdot\, |\, \cdot\, )_{T}$ is positive definite.
\eel
\proof Clearly, if $(G|G)_T=0$, both terms on the r.h.s. of (\ref{scalar-product}) must vanish. Suppose that
\beqa
\lan G_1, \mu^{-1/2}(T_1^*T_1)\mu^{-1/2} G_1\ran=\|T_1 \mu^{-1/2} G_1\|_2^2=0.
\eeqa
Then,  the property below (\ref{T-def}) gives
\beqa
0=\lan  T_2 \mu^{1/2} G_1, T_1 \mu^{-1/2} G_1\ran=\lan G_1, G_1\ran=0.
\eeqa
The second term on the r.h.s. of (\ref{scalar-product}) is treated analogously. \qed\\
Of particular importance for us will be the scalar product $( \,\cdot\,|\,\cdot\,)$ corresponding to $T=\mrm{id}$.
Using it, we can write
\beqa
S_{T}(G,G')=( G | \wt{S}_T G'), \quad\textrm{where} \quad \wt{S}_T=\begin{pmatrix} 
 \mu^{1/2}T_1^*T_1 \mu^{-1/2}    & \i \mu \\
-\i\mu^{-1} &    \mu^{-1/2}T_2^*T_2    \mu^{1/2} 
\end{pmatrix}.
\eeqa
Now we state the criterion of Araki-Yamagami in a form adapted to our problem.
\bet\label{Araki-Yamagami} \emph{\cite{AY82}} Fix $r>0$. The representation $\pi_T$ is $\si$-weakly continuous on $\mfa(\mco_r)$  if and only if the following two conditions are satisfied:
\begin{enumerate}
\item There is $C_r > 0$ s.t.   $C_r^{-1}(G|G) \leq  (G|G)_T \leq  C_r(G|G) $ for all $G\in L_r$.

\item  $\wt{S}_T^{1/2}-\wt{S}^{1/2}$ is a Hilbert-Schmidt operator on the Hilbert space $(L_r^{\mrm{cpl}}, (\,\cdot\,|\,\cdot\,))$.
\end{enumerate}
\eet
It is easy to check that assumptions 1., 2. of Theorem~\ref{main-result} imply, respectively, conditions 1., 2., 
in Theorem~\ref{Araki-Yamagami}. As the case of condition 1. is obvious, we move on to condition 2.  By \cite[Appendix B]{Bu74},  it suffices to show that  
\beqa
\wt{S}_T-\wt{S}=\begin{pmatrix} 
 \mu^{1/2}(T_1^*T_1-1)\mu^{-1/2}    & 0 \\
0 &    \mu^{-1/2} (T_2^*T_2-1) \mu^{1/2} 
\end{pmatrix}
\eeqa
is trace class  on $(L_r^{\mrm{cpl}}, (\,\cdot\,|\,\cdot\,))$.  This latter property is implied by the trace class property on $L^2(\real^3)$ of operators $K_{1,r}, K_{2,r}$ of (\ref{K-operators}). This concludes the proof of Theorem~\ref{main-result}.
\section{Kraus-Polley-Reents infravacuum maps}\label{KPR-maps}
\setcounter{equation}{0}

In this section we apply Theorem~\ref{main-result} to prove local normality of infravacuum 
representations. To define them, we will use the decomposition $L^2(\real^3)=L^2(\real_+)\otimes L^2(S^2)$
corresponding to spherical coordinates, where the measure of the second factor is normalized to the area of the
sphere $S^2$.
\bed\label{infravacuum}  The  Kraus-Polley-Reents infravacuum  maps $T_j: \mcL_j\to L^2(\real^3)$, $j=1,2$, are defined as follows:

\begin{itemize}

\item We introduce  sequences $\eps_i:=2^{-(i-1)}\ka$ and $b_i:= \fr{1}{i}$ for $i=1,2,3\ldots$.

\item We define functions $\xi_i(|\vk|):=\fr{\chi_{[\eps_{i+1},\eps_i] }(|\vk|)}{|\vk|^{3/2}}\in L^2(\real_+)$ and their normalized  counterparts $\ti\xi_i(|\vk|):= \xi_i(|\vk|) / \|\xi_i\|_{L^2(\real_+)}$. 

\item 
We define the orthogonal projections $Q_i: L^2(\real^3)  \to L^2(\real^3)$
and $\tilde{Q}_i: L^2(S^2)\to L^2(S^2)$ given by
\begin{equation}\label{pmbQi}
Q_i = |\ti\xi_i\ran\lan \ti \xi_i| \otimes \tilde{Q}_i \quad \text{with} \quad \tilde{Q}_i := \sum_{0\leq  \ell \leq i} 
\sum_{m=-\ell}^\ell   | Y_{\ell m } \rangle \langle Y_{\ell m } |,
\end{equation}
where $Y_{\ell m}$ are the spherical harmonics. 

\item We introduce the complex-linear maps $T_j: \mcL_j\to L^2(\real^3)$, $j=1,2$, 
\begin{equation}\label{T1T2}
 T_1 := I + \operatorname*{s-lim}_{n \to \infty} \sum_{i =1}^n (b_i -1)Q_i, \quad 
 T_2 := I +\operatorname*{s-lim}_{n \to \infty} \sum_{i=1}^n \big(\frac{1}{b_i} - 1 \big)Q_i.
\end{equation}
These maps are well-defined by Lemma~\ref{smoothing} below. 
We will denote by $T_{1,n}, T_{2,n}$ the respective approximants.
\end{itemize}
\eed
This definition is fine-tuned in such a way that $\pi_T$ is an irreducible infravacuum representation w.r.t. the subgroup 
$S$ of coherent automorphisms as in Lemma~\ref{coherent-automorphisms} \cite{Ku98, CD18}. Thus we can focus on the problem of local normality.

The assumptions of Theorem~\ref{main-result} are formulated in terms of $T_j^*T_j$, $j=1,2$. 
They can be expressed as follows as quadratic forms on  $\mcL_j$:
\beqa
& &T_{1}^2=I+\sum_{i =1}^{\infty} (b_i^2 -1)Q_i, \quad
 T_{2}^2=I+\sum_{i =1}^{\infty} (b_i^{-2} -1)Q_i.
\eeqa
We note that
\beqa
\chi_{j,r} Q_i \chi_{j,r}^*= \chi_{j,r} \chi'_{j,r}  Q_i \chi'{}^*_{j,r} \chi_{j,r}^*,
\eeqa
where $\chi_r'$ is an approximate characteristic function of $O_r$ s.t. $\chi'_r\chi_r=\chi_r$. We can write
\beqa
Q_{i, j,r}:=\chi'_{j,r}  Q_i \chi'{}^*_{j,r}\1&=&\1\sum_{0\leq  \ell \leq i} 
\sum_{m=-\ell}^\ell  \chi'_{j,r} | \ti \xi_i\otimes  Y_{\ell m } \rangle \langle \ti \xi_i\otimes Y_{\ell m } | \chi'{}^*_{j,r}\non\\
\1&=&\1 \sum_{0\leq  \ell \leq i} 
\sum_{m=-\ell}^\ell  \|  {\xi}^{j,r}_{i, \ell m}  \|^2_2 \,  |\ti{\xi}^{j,r}_{i, \ell m} \ran \lan \ti{\xi}^{j,r}_{i, \ell m}|, \label{Q}
\eeqa
where   ${\xi}^{j,r}_{i, \ell m}:=\chi'_{j,r} (\ti \xi_i\otimes  Y_{\ell m })$,     
$\ti{\xi}^{j,r}_{i, \ell m}:=\fr{ {\xi}^{j,r}_{i, \ell m} }{\| {\xi}^{j,r}_{i, \ell m}     \|_2 }$ and, by Lemma~\ref{smoothing},
\beqa
\|  {\xi}^{j,r}_{i, \ell m}  \|^2_2\leq C_r \eps_i^2, \quad  \|Q_{i, j,r}\|\leq C_r (i+1)^2\eps_i^2. \label{Q-smoothing}
\eeqa
Due to these estimates, the following operators
\beqa
(T_{1}^2)_r:=I+\sum_{i =1}^{\infty} (b_i^2 -1)Q_{i,j,r}, \quad (T_{2}^2)_r:=I+\sum_{i =1}^{\infty} (b_i^{-2} -1)Q_{i,j,r}
\eeqa
are bounded. As they satisfy $\chi_{j,r} T^2_j \chi_{j,r}^*=\chi_{j,r} (T^2_{j})_r \chi_{j,r}^*$, we immediately 
obtain the second inequality in assumption 1. of  Theorem~\ref{main-result}. 

As for the first inequality, the case of $T_{2}^2$ is immediate: Since $(b_i^{-2} -1)\geq 0$, we can write 
\beqa
\chi_{2,r} \chi^*_{2,r}\leq \chi_{2,r} (1+\sum_{i =1}^{\infty} (b_i^{-2} -1)Q_i)\chi^*_{2,r}=  \chi_{2,r}T_{2}^2\chi_{2,r}.
\eeqa 
In the case of $T_{1}^2$ we have $(b_i^{2} -1)\leq 0$, thus the above argument does not apply. Instead, we proceed as follows: Fix some $N\in \nat$ and write
\beqa
\chi_{1,r}  T_{1}^2  \chi^*_{1,r}=\chi_{1,r} T_{1,N}^2\chi^*_{1,r}+\chi_{1,r} \sum_{i =N+1}^{\infty} (b_i^2 -1)Q_{i,j,r}\chi^*_{1,r}, \label{T-one-lower-bound}
\eeqa 
where $T_{1,N}$ is the approximant as defined below (\ref{T1T2}). We note that the spectrum of $T_{1,N}^2$
can be read off directly from its definition. Thus we can write 
\beqa
T_{1,N}^2\geq  \inf\mrm{sp}(T_{1,N}^2)I=  b_{N}^2I=N^{-2} I.
\eeqa
On the other hand
\beqa
\|\sum_{i =N+1}^{\infty} (b_i^2 -1)Q_{i,j,r}\|\leq \sum_{i =N+1}^{\infty} |b_i^2 -1|\| Q_{i,j,r}\|\leq 
C_r\sum_{i =N+1}^{\infty}  (i+1)^2\eps_i^2\leq C_r' 2^{-N/2}.
\eeqa
Coming back to (\ref{T-one-lower-bound}), 
\beqa
\chi_{1,r}  T_{1}^2  \chi^*_{1,r} \geq  \big(N^{-2}- C_r' 2^{-N/2} \big) \chi_{1,r}\chi^*_{1,r}.
\eeqa
Now for any given constant $C_r'$ we can choose $N$ s.t.   $N^{-2}- C_r' 2^{-N/2}>0$, which concludes our
verification of assumption 1. of Theorem~\ref{main-result}.

To verify assumption 2, we define $\chi_{1,r}^{0}:=\chi_{r} \mu^{-1/2}$, $\chi_{2,r}^{0}:=\chi_r \mu^{1/2}$. 
Analogously as in (\ref{Q}), we write
\beqa
Q^0_{i, j,r}:=\chi^0_{j,r}  Q_i (\chi^0_{j,r})^* =  \sum_{0\leq  \ell \leq i} 
\sum_{m=-\ell}^\ell  \|  {\xi}^{j,r,0}_{i, \ell m}  \|^2_2 \,  |\ti{\xi}^{j,r,0}_{i, \ell m} \ran \lan \ti{\xi}^{j,r,0}_{i, \ell m}|.
\eeqa
By items (\ref{smooth-three}), (\ref{smooth-four}) in Lemma~\ref{smoothing}, the estimates of  (\ref{Q-smoothing})
hold also in this case, that is,
\beqa
\|  \ti{\xi}^{j,r,0}_{i, \ell m}  \|^2_2\leq C_r \eps_i^2, \quad  \|Q_{i, j,r}^0\|\leq C_r (i+1)^2\eps_i^2.
\eeqa
Thus the operator
\beqa
K_{1,r}:= \chi_{1,r}^0(T_1^2-1)(\chi_{1,r}^0)^*=  \sum_{i=1}^{\infty}  (b_i^2 -1) Q_{i, j,r}^0
\eeqa
is obviously trace-class on $L^2(\real^3)$ and the same is true for $K_{2,r}$. 

We  summarize our considerations in this paper as follows:
\bet Let $T$ be the  KPR  map of Definition~\ref{infravacuum}. Then the representation $\pi_T$
is irreducible, locally normal and has the infravacuum property w.r.t. $S$ defined above~(\ref{infravacuum-two}). 
Thus the automorphism $\al_T$, associated with $\pi_T$ via (\ref{Takesaki-alpha}), belongs to
the relative normalizer $N_{G}(R,S)$, where $G=\mrm{Aut}(\mfa)$ and $R=G_{x_0}$ is the stabilizer 
of the vacuum sector.  
 \eet
\appendix
\section{Technical lemmas} \label{Appendix}
\setcounter{equation}{0}
\bel\label{smoothing} There hold the bounds
\beqa
\|\mu^{1/2} \chi_r \mu^{-1/2} (\ti\xi_i\otimes  Y_{\ell m})\|_2 \1&\leq&\1 C_r \eps_i, \label{smooth-one}\\
\| \chi_r \mu^{-1/2} (\ti\xi_i\otimes  Y_{\ell m})\|_2 \1&\leq&\1 C_r \eps_i, \label{smooth-three}\\
\|\mu^{-1/2} \chi_r \mu^{1/2} (\ti\xi_i\otimes  Y_{\ell m})\|_2 \1&\leq&\1 C_r \eps_i^2, \label{smooth-two}\\
\| \chi_r \mu^{1/2} (\ti\xi_i\otimes  Y_{\ell m})\|_2 \1&\leq&\1 C_r \eps_i^2, \label{smooth-four}
\eeqa
for some $C_r$ independent of $i, \ell,m$.
\eel
\proof Starting with (\ref{smooth-one}), we can write
\beqa
\|\mu^{1/2} \chi_r \mu^{-1/2} (\ti\xi_i\otimes  Y_{\ell m})\|_2 \1 &=&\1 \|\mu^{1/2} \chi_r \mu^{-1/2} \chi_{[\eps_{i+1}, \eps_{i} ]}(\ti\xi_i\otimes  
Y_{\ell m})\|_2\non\\
\1&\leq &\1  \|\mu^{1/2} \chi_r \mu^{-1/2} \chi_{[\eps_{i+1}, \eps_{i} ]}\|,
\eeqa
where $\chi_{[\eps_{i+1}, \eps_{i} ]}$ is the operator of multiplication by the sharp characteristic function 
$|k|\mapsto \chi_{[\eps_{i+1}, \eps_{i} ]}(|k|)$ of $[\eps_{i+1}, \eps_{i} ]$ and
in the last line the operator norm is understood. We will estimate this norm using the Schur lemma \cite[Section B6]{DG}: If $A$ is an operator and $a$ its kernel, then $\|A\|\leq (CC')^{1/2}$  provided that
\beqa
\sup_k\int |a(k,k')| dk' \leq C\quad  \textrm{and} \quad   \sup_{k'}\int |a(k,k')| dk \leq C'.  \label{two-integrals}
\eeqa 
In our case $a(k,k')=(2\pi)^{-3/2} |k|^{1/2}\hat{\chi}_r(k-k')|k'|^{-1/2} \chi_{[\eps_{i+1}, \eps_{i} ]}(|k'|)$. We have 
\beqa
\int |a(k,k')| dk' \1&\leq&\1 
(2\pi)^{-3/2}\fr{ 1}{\eps_{i+1}^{1/2} }\int_{\real^3} (|k-k'|^{1/2}+1) |\hat{\chi}_r(k-k')|  \chi_{[\eps_{i+1}, \eps_{i} ]}(|k'|)dk'\non\\
\1&\leq&\1 \fr{c}{\eps_{i+1}^{1/2}} \int_{\eps_{i+1}}^{\eps_i} |k|^2d|k| \leq c' \eps_i^{5/2}.
\eeqa
Now the second integral in (\ref{two-integrals}) can be estimated as follows
\beqa
\int |a(k,k')| dk \1&\leq&\1  \fr{c}{\eps_{i+1}^{1/2} }\int_{\real^3} (|k-k'|^{1/2}+1) |\hat{\chi}_r(k-k')|  \chi_{[\eps_{i+1}, \eps_{i} ]}(|k'|)dk \leq   \fr{c'}{\eps_{i}^{1/2}}. \quad\quad
\eeqa
Thus we have $\|\mu^{1/2} \chi_r \mu^{-1/2} \chi_{[\eps_{i+1}, \eps_{i} ]}\|\leq  c'' \eps_i$
which gives (\ref{smooth-one}).  Estimate (\ref{smooth-three}) is an immediate consequence,
since
\beqa
\| \chi_r \mu^{-1/2} (\ti\xi_i\otimes  Y_{\ell m})\|_2 \1&=&\1 \|\chi'_r\mu^{-1/2} \mu^{1/2} \chi_r \mu^{-1/2} (\ti\xi_i\otimes  Y_{\ell m})\|_2\non\\
\1&\leq&\1 \|\chi'_r\mu^{-1/2} \| C_r \eps_i, 
\eeqa
and $\|\chi'_r\mu^{-1/2} \|<\infty$  by    Lemma~\ref{standard}  below. 

Let us move on to (\ref{smooth-two}). In this case we write
\beqa
\|\mu^{-1/2} \chi_r \mu^{1/2} (\ti\xi_i\otimes  Y_{\ell m})\|_2\leq \|\mu^{-1/2} \chi'_r\|\,  \|
\chi_r \mu^{1/2}\chi_{[\eps_{i+1}, \eps_{i} ]} \|. 
\eeqa
Now we estimate the norm of
$\chi_r \mu^{1/2}\chi_{[\eps_{i+1}, \eps_{i} ]}$ using the Schur lemma.  The 
kernel has now the form $a'(k,k')= (2\pi)^{-3/2} \hat{\chi}_r(k-k') |k'|^{1/2}\chi_{[\eps_{i+1}, \eps_{i} ]}(k')$.
We immediately see that
\beqa
\int |a(k,k')| dk' \leq c \eps_i^{7/2}, \quad   \int |a(k,k')| dk\leq c\eps_i^{1/2},  
\eeqa
which gives (\ref{smooth-two}) and (\ref{smooth-four}). \qed
\bel\label{standard} The operators $\chi_r \mu^{-1/2}$ and $\mu^{1/2}\chi_r \mu^{-1/2}$  extend from $D(\real^3;\complex)$ to
 bounded operators on $L^2(\real^3)$.
\eel
\proof We refer to \cite[Lemma~3.2]{Dy08} for boundedness of  $\chi_r \mu^{-1/2}$. 
As for the second operator, its kernel satisfies
\beqa
|a(k,k')|\1& =&\1 (2\pi)^{-3/2}  |k|^{1/2} |\hat\chi_r(k-k')| |k'|^{-1/2}\non\\
\1&\leq&\1(2\pi)^{-3/2}\big( |k-k'|^{1/2}  |\hat\chi_r(k-k')| |k'|^{-1/2}\!+\!|\hat\chi_r(k-k')|\big). \quad
\eeqa
Hence, for any $G,G'\in D(\real^3;\complex)$ we can write
\beqa
|\lan G, \mu^{1/2}\chi_r \mu^{-1/2}G'\ran|\leq   c\big(\lan |G|,    \wt{\chi}_r \mu^{-1/2} |G'|\ran       + \|G\|_2\|G'\|_2\big),
\eeqa  
where $\wt{\chi}_r$ acts by convolution with the rapidly decaying function $k\mapsto  |k|^{1/2}|\hat\chi_r(k)|$.
Now boundedness of $\wt{\chi}_r \mu^{-1/2}$ follows by analogous arguments as boundedness of $\chi_r \mu^{-1/2}$.
\qed

\end{document}